\documentclass[aps,nofootinbib,preprintnumbers,twocolumn,superscriptaddress,prb,showpacs]{revtex4}

\usepackage{amsmath}
\usepackage{amssymb}
\usepackage{amsfonts}
\usepackage{color,array}
\usepackage{dsfont}
\usepackage{slashed}
\usepackage{graphicx}
\usepackage{graphics}
\usepackage{epsfig}
\usepackage{bbm,bm}
\usepackage{psfrag}
\usepackage{hyperref}
\usepackage{ulem}
\usepackage{url}
\newcommand{\be}{\begin{equation}}
\newcommand{\ee}{\end{equation}}
\newcommand{\ben}{\begin{eqnarray}}
\newcommand{\een}{\end{eqnarray}}

\begin{document}

\title{Multicritical behavior in models with two competing order parameters}
\date{\today}

\author{Astrid Eichhorn}
\email{aeichhorn@perimeterinstitute.ca}
\affiliation{\mbox{\it Perimeter Institute for Theoretical Physics, 31 Caroline Street N, Waterloo, N2L 2Y5, Ontario, Canada}
}

\author{David Mesterh\'azy}
\email{d.mesterhazy@thphys.uni-heidelberg.de}
\affiliation{\mbox{\it Institut f\"ur Theoretische Physik,
Universit\"at Heidelberg,}
\mbox{\it D-69120 Heidelberg, Germany}
}

\author{Michael M. Scherer}
\email{m.scherer@thphys.uni-heidelberg.de}
\affiliation{\mbox{\it 	Institut f\"ur Theoretische Physik,
Universit\"at Heidelberg,}
\mbox{\it D-69120 Heidelberg, Germany}
}

\pacs{05.10.Cc,05.70.Jk,64.60.ae,64.60.F-}



\begin{abstract}
We employ the nonperturbative functional Renormalization Group to study models with an $O(N_1) \oplus O(N_2)$ symmetry.
Here, different fixed points exist in three dimensions, corresponding to bicritical and tetracritical
behavior induced by the competition of two order parameters. We discuss the critical behavior of the symmetry-enhanced isotropic, the decoupled and the biconical fixed point, and analyze their stability in the $N_1, N_2$ plane.
We study the fate of non-trivial fixed points during the transition from three to four dimensions, finding evidence for a triviality problem for coupled two-scalar models in high-energy physics. We also point out the possibility of non-canonical critical exponents at semi-Gaussian fixed points and show the emergence of Goldstone modes from discrete symmetries.
\end{abstract}

\maketitle

\section{Introduction}

Systems with competing order parameters and their corresponding symmetries play an important role in a variety of physical situations, where the competing orders entail multicritical points in the phase diagram \cite{FisherLiu,KosterlitzNelsonFisher, NelsonKosterlitzFisher,Aharony,Aharony2,Calabrese:2002bm,Moser2008}, for reviews see Refs.~\onlinecite{Aharony3,Vicari:2007ma, Pelissetto1, Pelissetto2}. Examples include anisotropic antiferromagnets in an external magnetic field \cite{Rohrer1975, Rohrer1977,KingRohrer,Oliveira, Butera,Ohgushi,beccera88,basten80}, high-$T_c$ superconductors \cite{Zhang1997, Zhang2} and possibly even Quantum Chromodynamics \cite{Schaefer:2004en,Fukushima:2013rx}. In the vicinity of a multicritical point, the phase diagram of such systems can be analyzed using a model of two coupled order parameter fields with $O(N_1) \oplus O(N_2)$ symmetry, where $N_1$ and $N_2$ depend on the physics in question. For instance, anisotropic antiferromagnets in a magnetic field along the easy axis have a phase diagram that can be described by a model with $N_1=1$ and $N_2=2$, corresponding to an antiferromagnetic phase and a spin flop phase, respectively, see Refs.~\onlinecite{NelsonFisher,Folkproc} for illustrations of the phase diagram.

\begin{figure}[!here]
\includegraphics[width=0.4\linewidth]{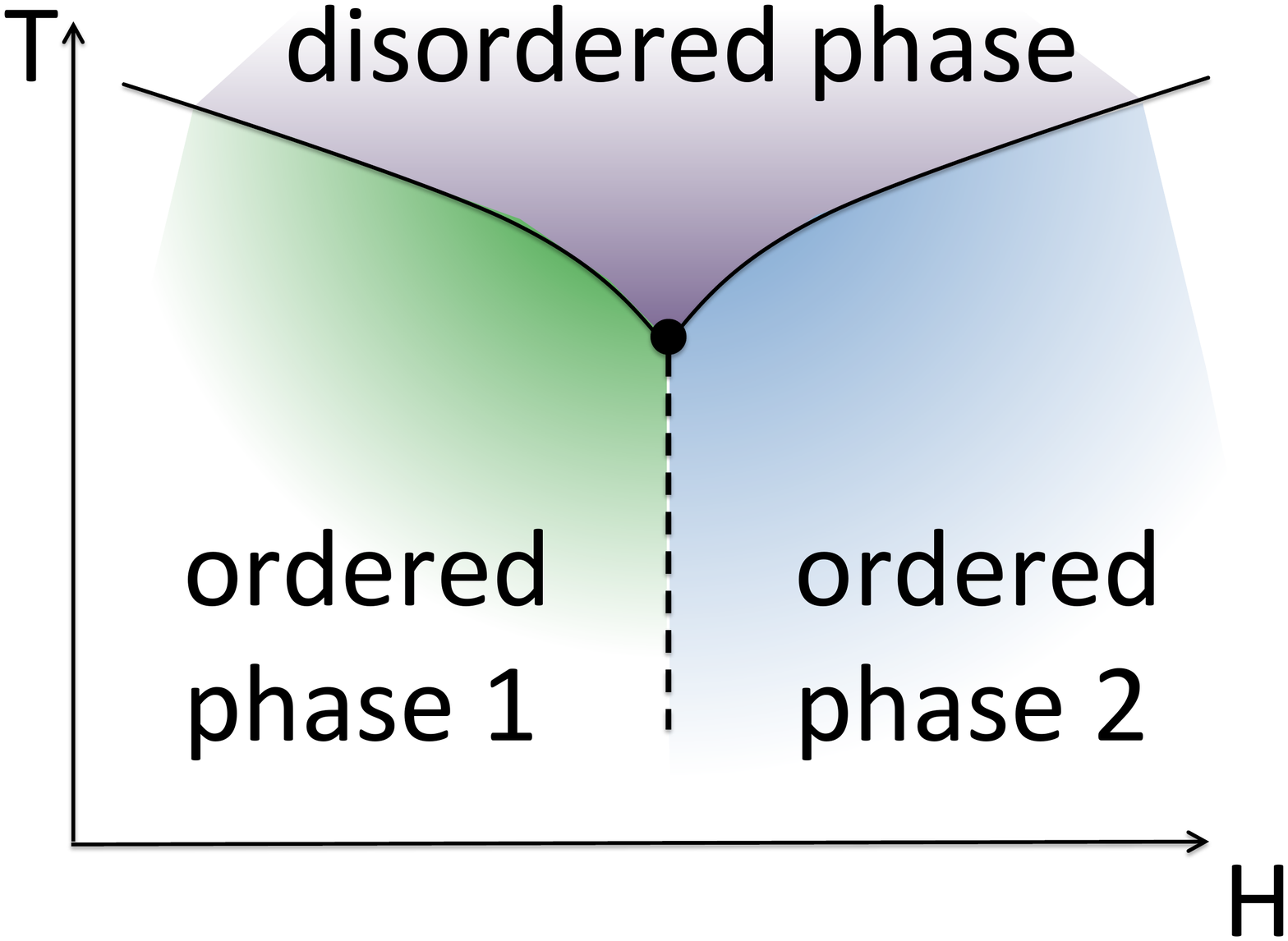}
\includegraphics[width=0.4\linewidth]{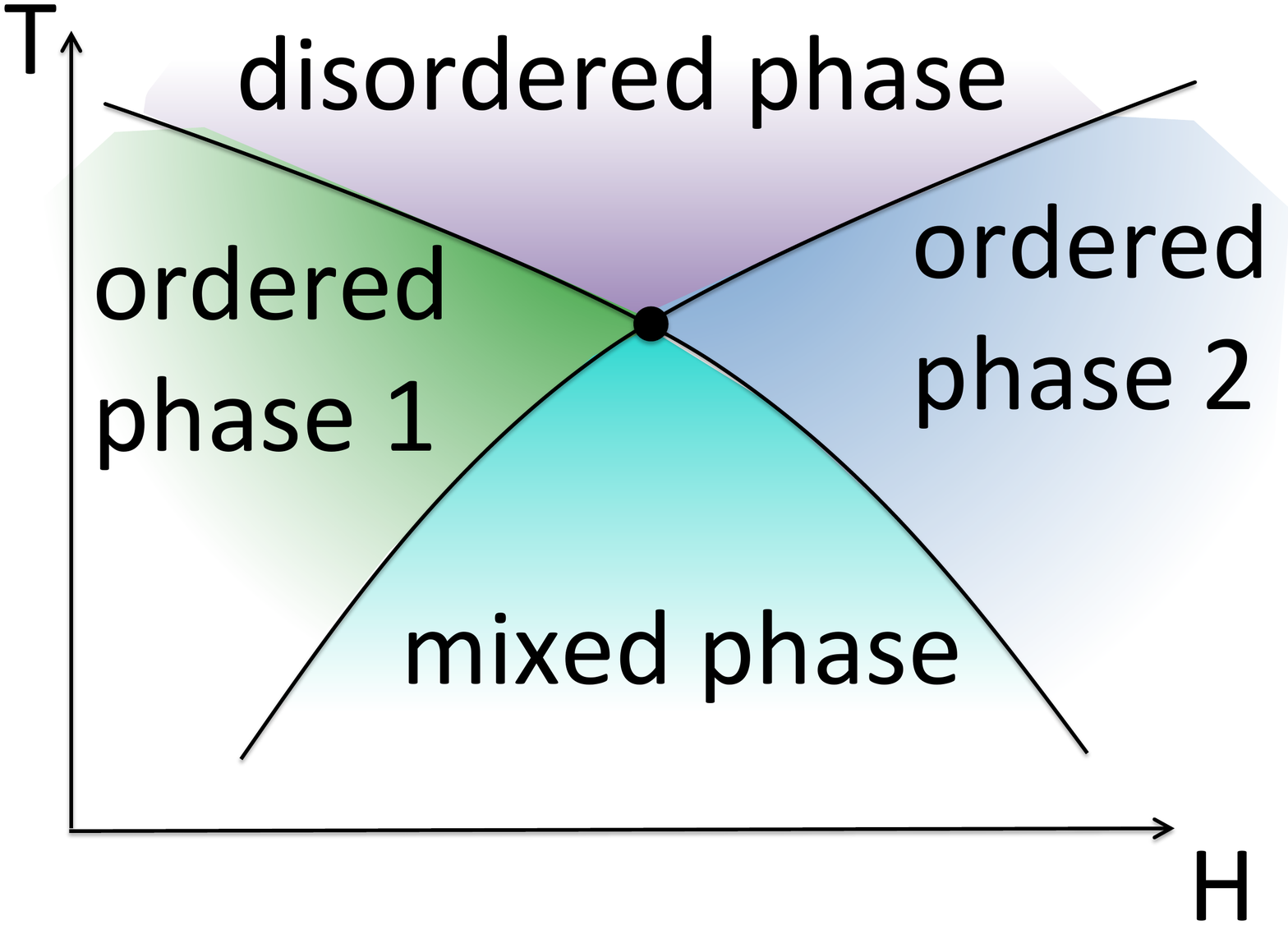}
\vspace{-0.35cm}
\caption{\label{biandtetra}(Color
 online) Here we plot schematic phase diagrams containing a bicritical point (left panel) and a tetracritical one (right panel), at which a mixed phase exists which is dominated by two condensing order parameters. Solid lines denote second order, and dashed lines first order phase transitions.}
\end{figure}

Models with competing order parameters were first investigated in Ref.~\onlinecite{FisherLiu}, where the existence of multicritical points, at which different phases meet, was pointed out: A bicritical point separates the two ordered phases and a phase of unbroken symmetry, see Fig.~\ref{biandtetra}. In contrast, four transition lines meet at a tetracritical point. The fourth phase is determined by two non-zero order parameters. In the case of superfluid helium, this phase has been termed supersolid  phase \cite{FisherLiu}, however, its experimental realization remains unclear, see, e.g., Refs. \onlinecite{Kim2004,Kim2012}.

As a function of $N_1$ and $N_2$, different phase diagrams are realized, with a bicritical point with symmetry enhancement $O(N_1) \oplus O(N_2) \rightarrow O(N_1+N_2)$ for $N_1=1=N_2$. A tetracritical point appears beyond critical values for $N_1$ and $N_2$, first calculated approximately in Refs.~\onlinecite{Aharony,KosterlitzNelsonFisher}. The determination of these critical values is possible with different methods, namely the $\epsilon$-expansion around $d=4-\epsilon$ dimensions \cite{NelsonKosterlitzFisher, KosterlitzNelsonFisher, Aharony, Calabrese:2002bm}, two-loop perturbative Renormalization Group methods \cite{Moser2008,Fedorenko1998}, as well as Monte Carlo simulations \cite{Caselle, Selke, Campostrini, Hasenbusch, Hu}. 

The nature of the multicritical point at $N_1=1$, $N_2=2$ is still under debate: Experiments indicate a bicritical point \cite{Rohrer1977, KingRohrer, Oliveira}, confirmed by early results from the $\epsilon$-expansion \cite{NelsonKosterlitzFisher, KosterlitzNelsonFisher, NelsonFisher}. Subsequent higher-order computations indicated an instability of the bicricital, symmetry enhanced fixed point \cite{Calabrese:2002bm}, see also Refs.~\onlinecite{Moser2008,Caselle}, implying a tetracritical nature of the point. In contrast, Monte Carlo simulations in Ref.~\onlinecite{Selke} found a bicritical point, while simulations in Ref.~\onlinecite{HasenbuschVicari} refuted these claims.

There are also several further interesting and open questions: The situation in $d=2$ remains to be fully understood, see, e.g., Ref. \onlinecite{Pelissetto3}. Also, the nature of the phase diagram close to the multicritical point depends on non-universal quantities specific to the material under consideration. These non-universal properties are more challenging to access theoretically. Furthermore, the existence of a bicritical point in certain regions of parameter space has not been studied in detail, yet. A nonperturbative method that provides access to fixed-point properties as well as non-universal quantities, also at finite temperature, is clearly indicated to tackle these questions.

It turns out that there is a wide class of matrix models, which can be reduced to a coupled theory of two distinct order parameters in a certain range of their parameter space, and which provide a more general context for these models\cite{Wegner:1979,Wegner:1979zz,Lee:1985zzc,Zirnbauer:1996zz,Brody:1981cx,Dyson:1962es,Osborn:1998qb,Shuryak:1992pi,Di Francesco:1993nw,Guhr:1997ve}.

Interestingly, the same models play a role in a rather different context, where the emphasis is not on competing order parameters:
In high-energy physics, a four-dimensional model with $N_1= 1=N_2$, realizing a $Z_2$ reflection symmetry for each scalar field, plays a role in hybrid models for inflation \cite{Linde:1993cn}, see Refs.~\onlinecite{Lyth, Brandenberger:2000as,Linde:2007fr} and references therein, and is also of interest as a toy model for a Higgs-inflaton coupling.

Here, we analyze these systems with the functional Renormalization Group (FRG)\cite{Wetterich:1993yh}, which allows us to analyze quantum and statistical field theories even away from the perturbative regime. This method provides a unified framework to access universal as well as non-universal properties. Thus, not only universal behavior in the vicinity of a second-order phase transition, but also physical properties away from the transition, as well as first-order phase transitions, can be studied in detail, see Refs.~\onlinecite{Braun:2007bx,Braun:2009gm}, for examples in the case of the QCD phase diagram.
The FRG is also applicable in any dimension, and does not require a small parameter to expand in. Instead our approximation consists of a particular truncation of the space of operators that drive the RG flow. Further, within perturbative RG approaches non-trivial resummation techniques are required in order to obtain reliable quantitative results, e.g., for critical exponents in three-dimensional $O(N)$ models. For the FRG method no such resummation is required, see Ref.~\onlinecite{delamotte2004, delamotteA, delamotteB} for a detailed discussion of this matter.
In many applications fermions also play an important role, and can be included straightforwardly with the FRG, even within the chiral limit, see for examples Refs.~\onlinecite{Jungnickel:1995fp,Rosa:2000ju,Hofling:2002hj,Strodthoff:2011tz,Janssen:2012pq,boett2012,mesterhazy2012,Scherer:2012tc}.
Unlike lattice field theory the FRG is a continuum method, and our results are therefore unaffected by either finite-volume or discretization artefacts. Results from the perturbative RG, Monte Carlo simulations and the FRG thus complement each other, and taken together provide us with the possibility to understand physical systems in great qualitative as well as quantitative detail. Here, we will focus on the analysis of multicritical points in $O(N_1) \oplus O(N_2)$ models in $3 \leq d \leq 4$.

Our main results consist in a confirmation of the position of the border between bicritical and tetracritical behavior in the $N_1, N_2$ plane as discussed in Refs. \onlinecite{Calabrese:2002bm, Moser2008} with the method of the nonperturbative functional Renormalization Group that we newly apply to these systems. We infer that the $N_1=1, N_2=2$ model shows tetracritical behavior with critical exponents corresponding to that of the so-called biconical fixed point. As a new result, we show how the critical behavior of the system is determined by collisions of fixed points in the space of couplings: As a function of $N_1, N_2$, the fixed points move through the space of coupling, exchanging stability properties as they collide. This observation could explain why the critical exponents describing the system in the $N_1=1, N_2=2$ case are close to the Heisenberg universality class. Further, we report the discovery of a new fixed point. Unlike the other fixed points, it does not show an enhancement of the symmetry group. It exists in a particular region of parameter space for these models, where previous analysis have not seen fixed points. In the case that this new fixed point persists beyond the approximation in our work, it would imply bicritical behavior for models in this region of parameter space. 
Finally, we turn to $d=4$ dimensions, and analyze implications for high-energy physics. We discuss the existence of noncanonical scaling behavior at fixed points which show vanishing fixed-point values for some couplings. Similar semi-interacting fixed points could be of interest for a UV completion of the Standard Model in the context of quantum gravity. In this case, similar noncanonical scaling behavior could be expected even for asymptotically free couplings.
We also follow the fixed points that exist in $d=3$ towards four dimensions and find that only the noninteracting fixed point can exist in $d=4$. This implies a triviality problem for models with coupled scalar degrees of freedom, with potential interest for Higgs models and inflaton models. Finally, we discuss how the enhancement of discrete symmetries to a continuous symmetry in the $N_1 =1=N_2$ case can lead to the emergence of Goldstone bosons from a model with discrete symmetries at the microscopic level. We explicitly follow an RG trajectory, starting from a microscopic action with discrete symmetry, and ending at a symmetry-enhanced fixed point. The masses of the two modes, starting out equal in the UV, differ in the IR, where one of them goes to zero.

This paper is structured as follows: In Sec.~\ref{model}, we define our model and discuss the possible phase diagrams and the nature of the multicritical point. 
The functional Renormalization Group as a calculational tool to access these models is explained in Sec.~\ref{method}. We specify our truncation and derive a flow equation for the effective potential in Sec.~\ref{truncation}. A connection to matrix models, that explains how to map  a subclass of matrix models onto our vector model is established in Sec.~\ref{matrixmodels}.
We present our results in Sec.~\ref{results}, first focussing on fixed points discussed in the literature, where we confirm results obtained with other methods. We then find additional fixed points in a disconnected part of parameter space, which we analyze in detail. Finally, we discuss the relevance of our results for high-energy physics. Here, a semi-Gau\ss{}ian fixed point i.e., a fixed point at which a subset of the couplings vanishes, defines an interesting new universality class, that shows that canonical scaling should not be expected for vanishing couplings at semi-Gau\ss{}ian fixed points. Similar behavior could be relevant for a UV completion of the Standard Model coupled to gravity. We then focus on the emergence of Goldstone bosons from a model with discrete symmetries. This mechanism relies on the possibility of enhancing the discrete symmetry to a continuous one at a fixed point of the Renormalization Group flow. Finally we present the continuation of the new fixed points towards $d=4$ dimensions and confirm the triviality of two-scalar models.
We conclude in Sec.~\ref{conclusions}.

\section{Model}\label{model}

Here, we study models that are composed of two bosonic sectors with $O(N_1)$ and $O(N_2)$-symmetry, respectively.
Non-trivial interaction terms which couple the two sectors are compatible with this requirement, and will be responsible for considerably more interesting physics than in the simpler case of a bosonic model with only one sector.
For these models the Landau-Ginzburg-Wilson Functional reads
\begin{eqnarray}\label{eq:on1on2}
\mathcal{H}&=&\int d^{d}x\left[\frac{1}{2}(\partial_\mu \phi)^2+\frac{1}{2}(\partial_\mu \chi)^2+\frac{1}{2}r_{\phi}\phi^2+\frac{1}{2}r_{\chi}\chi^2\right.\nonumber\\
&&\quad\quad\quad+\left.\frac{u_{\phi}}{4!}\phi^4+\frac{u_{\chi}}{4!}\chi^4+\frac{u_{\phi\chi}}{3\cdot4}\phi^2\chi^2\right]\,,
\end{eqnarray}
where $\phi=(\phi_1,\phi_2,...,\phi_{N_1})$ and $\chi=(\chi_1,\chi_2,...,\chi_{N_2})$ are $N_1$-component and $N_2$-component fields, respectively. The functional in Eq. \eqref{eq:on1on2} is symmetric under $O(N_1) \oplus O(N_2)$ transformations. Naturally, this model features multicriticality when the critical lines of the two different order parameters intersect. A variety of fixed points appears in those theories and it requires a stability analysis of the RG flow in the vicinity of the fixed points to find out which one governs the critical behavior. The number of relevant eigendirections, i.e., the number of negative eigenvalues of the stability matrix determines the number of parameters that have to be fine-tuned in order to approach the critical point. The infrared (IR) stable fixed point is the one with the lowest number of relevant directions. 

The bicritical versus tetracritical nature of the multicritical point, see Fig.~\ref{biandtetra}, depends on the sign of
\be
\Delta = u_{\phi} u_{\chi}- u_{\phi \chi}^2\,,
\ee
at the fixed point. For $\Delta>0$ the fixed point is tetracritical, whereas it becomes bicritical for $\Delta<0$, cf. Ref.~\onlinecite{FisherLiu}. In the first case, a phase 1 of unbroken symmetry, a phase 2 with broken $O(N_1)$ and unbroken $O(N_2)$, a phase 3 with unbroken $O(N_2)$ and broken $O(N_1)$, and a phase 4 with broken $O(N_1) \oplus O(N_2)$ meet at the multicritical point.
This can be derived directly from a consideration of the Gibbs free energy\cite{FisherLiu}, where it becomes clear that the phase with both symmetries broken can only exist for $\Delta>0$.

A special subspace of our theory space for which an additional symmetry $\phi \leftrightarrow \chi$ holds has been examined in Ref.~\onlinecite{Bornholdt:1994rf} for $N_1=1=N_2$ and at finite temperature in Ref.~\onlinecite{Bornholdt:1995rn}, with a special focus on $d=3$ and $d=4$. We will keep the dimensionality general in the following and later specialize to $d=3$ and $d=4$.

\section{Method}\label{method}

In the following we employ the nonperturbative functional Renormalization 
Group (FRG) to evaluate the generating functional (for reviews see, e.g., Refs. \onlinecite{Berges:2000ew,Polonyi:2001se,Pawlowski:2005xe,Gies:2006wv,Delamotte:2007pf,Rosten:2010vm,metzner2011}). This method allows to successively integrate out statistical (and quantum) fluctuations following the Kadanoff-Wilson picture of momentum shell integration\cite{Kadanoff:1966wm,Wilson:1971bg,Wegner:1972ih,Wegner:1976bn,Wilson:1973jj} in a Euclidean setting. For this description we start with the functional integral representation of the partition function $Z=\int_{\Lambda} \mathcal{D}\varphi\, e^{- S[\varphi]}$ with the microscopic action $S[\varphi]$, where $\Lambda$ is a UV cutoff. The flowing action is defined as a modified Legendre transform of the infrared regularized Schwinger functional $W_{k}[J]$, i.e.
\begin{eqnarray}
\Gamma_k[\Phi] &=& \sup_{J} \left\{ \int d^{d}x \, J(x) \Phi(x) - W_{k}[J] \right\} \nonumber\\ 
&& -\: \frac{1}{2} \int\!\! \frac{d^{d}p}{(2\pi)^{d}} \Phi(-p)R_k( p )\Phi ( p ),
\end{eqnarray}
where $\Phi = \langle \varphi \rangle_{J}$, and
\begin{equation}
W_{k}[J] = \ln \int_{\Lambda} \mathcal{D}\varphi\, e^{- S[\varphi]- \frac{1}{2} \int\!\! \frac{d^{d}p}{(2\pi)^{d}} \varphi(-p) R_k( p )\varphi ( p ) } ~.
\end{equation}
 $k$ is an infrared momentum scale. $J$ denotes a source-term. The function $R_k=R_k(q)$ acts as a masslike, $k$-dependent regulator suppressing infrared modes below the RG scale $k$. Up to the requirements that $R_k(q) \rightarrow \infty$ for $k \rightarrow \Lambda \rightarrow \infty$, $R_k(q) \approx k^2$ for $\frac{|q|}{k} \rightarrow 0$, and $R_k(q) \rightarrow 0$ for $\frac{k}{|q|} \rightarrow 0$ it can be chosen freely. The flowing action $\Gamma_k$ then contains the effect of fluctuations above the momentum scale $k$ only, and connects the microscopic action $S$ for $k \rightarrow \Lambda$ to the full effective action $\Gamma$ in the infrared. The latter is the generating functional of the one-particle-irreducible (1PI) correlation functions allowing to access the macroscopic or thermodynamic properties of the system under consideration.
The FRG then provides a functional differential equation for the flowing action $\Gamma_k$ whose scale-dependence is governed by the Wetterich equation \cite{Wetterich:1993yh},
\begin{equation}\label{eq:flowequwett}
 \partial_t \Gamma_k[\Phi]=\frac{1}{2}\mathrm{Tr}\left[ 
\left(\Gamma_k^{(2)}[\Phi]+R_k \right)^{-1} \partial_t R_k\right] \,,
\end{equation}
with $\partial_t = k \partial_k$. Here, the field $\Phi$ collects all bosonic degrees of freedom of a given model, and $\Gamma_k^{(2)}[\Phi]$ denotes the second functional derivative
of $\Gamma_k$
\begin{equation}
 \left( \Gamma_k^{(2)}[\Phi]\right)_{ij}(p_1,p_2)=\frac{\delta^{2}}{\delta\Phi_{i}(-p_1) \delta\Phi_{j}(p_2)} \Gamma_k[\Phi]\,,
\end{equation} 
where $p_i$ denote momenta. The Tr operation involves a summation over internal indices as well as a loop momentum integration. With these conditions, the solution to Eq. (\ref{eq:flowequwett}) provides an RG trajectory, interpolating between the microscopic action $\Gamma_\Lambda$ at the ultraviolet scale $\Lambda$ and the full effective action $\Gamma=\Gamma_{k\to 0}$.  

One of the main technical advantages of Eq. \eqref{eq:flowequwett} is its one-loop form, written as the trace over the full propagator, with the regulator insertion  $\partial_t R_k$ in the loop. It is crucial to stress that the method is nonperturbative and thus also yields higher terms in a perturbative expansion, see, e.g., Ref. \onlinecite{Reuter:1993kw}, since it depends on the full, field- and momentum-dependent propagator, and not just on the perturbative propagator.

An expansion of the flowing action functional $\Gamma_k$ in terms of a suitable
basis of momentum-dependent monomials in $\Phi$

\be
\Gamma_k = \sum_i \bar{g}_i(k) \mathcal{O}_i(\partial, \Phi)\,,
\ee
turns Eq. (\ref{eq:flowequwett}) into an infinite set of coupled differential equations for the expansion coefficients $\bar{g}_i(k)$, i.e., the running couplings, in terms of $\beta$ functions. A suitable expansion scheme should include the physically important degrees of freedom of a given problem at all scales under consideration and respect the symmetries of the system. Reducing the expansion to a tractable (and typically finite) subset of running couplings defines a truncation.
Crucially, the success of a chosen truncation does not necessarily rely on the existence of a small expansion parameter. While perturbative results can be reproduced straightforwardly with the Wetterich equation, its regime of validity goes beyond perturbation theory, and allows to access nonperturbative physics.
To devise a truncation that yields quantitatively good results only requires that the neglected operators do not couple too strongly into the flow of the included operators. The quality of a truncation can be tested by the convergence of the results under systematical extensions of a given truncation scheme, by a study of its regulator dependence and, of 
course, by comparison to well-known limiting cases as well as complementary methods. Convergence of the FRG flow can be improved by the choice an optimized regulator function\cite{Litim:2002cf,Litim:2001up,Pawlowski:2005xe}.


In the following, we will be interested in fixed-point solutions, corresponding to scale-free points, as these allow to evaluate the scaling behavior of physical quantities close to a second-order phase transition.
To that end we study the RG flow of the dimensionless couplings
\be
g_i(k) = \bar{g}_i(k) k^{-d_{g_i}},
\ee
where $d_{g_i}$ is the canonical dimensionality of the coupling. The reason for this is that a fixed point, which is a scale-free point, is hard to identify when dimensionful couplings are present, since each dimensionful coupling corresponds to a scale.
The RG flow of the couplings is given in terms of $\beta$ functions
\be
\beta_{g_i}(\{g_n\})= \partial_t g_i(k)= -d_{g_i} g_i(k) + f(\{g_n\}),
\ee
which are functions of the dimensionless couplings, and do not explicitly depend on the RG scale $k$. The first term reflects the scale-dependence due to the canonical dimensionality, whereas the second term carries the quantum/statistical corrections to the scaling.

To determine the critical exponents, which enter the scaling of observable quantities in the vicinity of a second-order phase transition, we linearize the flow around the fixed point $\{g_{n\,\ast}\}$, satisfying
\be
\beta_{g_i}(\{g_{n\,\ast}\})= 0 \,,
\ee
 and we find
\be
\beta_{g_i}(\{g_n\})= \sum_j\frac{\partial \beta_{g_i}}{\partial g_j} \Big|_{g_n = g_{n\, \ast}}\left(g_j - g_{j\, \ast} \right) +...\,.\label{linflow}
\ee
The solution to this linearized equation is given by
\be
g_{i}(k) = g_{i\, \ast} + \sum_I C_I V^{I}_i  \left(\frac{k}{k_0}\right)^{- \theta_I},
\ee
where 
\be
- \frac{\partial \beta_{g_i}}{\partial g_j} \Big|_{g_n = g_{n\, \ast}} V_I = \theta_I V_I.
\ee
Herein, $C_I$ is a constant of integration and $k_0$ is a reference scale. The $V_I$ are the eigenvectors and $-\theta_I$ the eigenvalues of the stability matrix, defined by \eqref{linflow}. The critical exponents $\theta_I$ can be complex, in which case the real part is decisive for the stability properties of the fixed point \cite{Kadanoff:2011aj}. 
In order to approach the fixed point in the IR, observe that $C_I$ is arbitrary for irrelevant directions where $\theta_I <0$. On the other hand, a relevant direction with $\theta_I>0$ corresponds to a parameter that needs to be tuned in order to ensure that the fixed point is reached in the IR. Accordingly, relevant directions correspond to quantitites that need to be adjusted experimentally (e.g., the temperature) in order to reach a second-order phase transition. We conclude that the fewer relevant directions a fixed point has, the more likely it plays a role in a realistic physical system, where only a small number of quantities is accessible to experimental tuning.

At a non-interacting fixed point, the $\theta_I$ equal the canonical dimensionality $d_{g_i}$, whereas non-trivial contributions are present at an interacting fixed point.

\subsection{Truncation and flow of the effective potential}\label{truncation}

To investigate models with two order parameter fields,
we consider a truncation of the form
\begin{equation}
\label{eq:AnsatzEffectiveAction}
\Gamma_k=\int d^{d}x\, \Big[ \frac{Z_{\phi\, k}}{2} \left(\partial_{\mu} \phi\right)^2+ \frac{Z_{\chi\, k}}{2} \left(\partial_{\mu} \chi\right)^2 + U_k(\phi, \chi)\Big]\,,
\end{equation}
with uniform scale-dependent wave-function renormalizations $Z_{\phi\, k}$ and $Z_{\chi\, k}$ and an effective potential $U_k(\bar\rho_{\phi}, \bar\rho_{\chi})$, where $\bar\rho_{\phi} =\frac{\phi^2}{2}$ and $\bar\rho_{\chi}= \frac{\chi^2}{2}$. These are the first terms in a derivative expansion of the effective action, i.e., a local potential approximation (LPA). We neglect further terms with a more complicated momentum- and field-dependence, as well as a distinction between Goldstone and massive modes in the anomalous dimension \cite{Berges:2000ew}. The scale derivatives of the wave function renormalizations are given by the anomalous dimensions
\be
\eta_{\phi} = -\partial_t {\rm ln}Z_{\phi\, k}, \quad\eta_{\chi} = -\partial_t {\rm ln}Z_{\chi\, k}. 
\ee
Using an optimized regulator \cite{Litim:2002cf,Litim:2001up,Pawlowski:2005xe} of the form $R_{k,\phi/\chi} (p) = Z_{\phi/\chi\, k}(k^2-p^2) \theta(k^2-p^2)$,
we can derive an equation for the dimensionless effective potential $u_k= k^{-d} U_k(\bar\rho_{\phi}, \bar\rho_{\chi})$ which is a scale-dependent quantity. 
As we are interested in a quantitative determination of critical exponents, the use of the optimized shape function is advisable \cite{Litim:2002cf}.
The effective potential $u_k$ is a function of the dimensionless renormalized field variables $\rho_\phi=Z_{\phi\, k} k^{2-d}\bar\rho_\phi,\quad \rho_\chi=Z_{\chi\, k} k^{2-d}\bar\rho_\chi$.  Its flow can be derived from \eqref{eq:flowequwett} and be written in compact form using threshold functions,
\ben
\partial_t u_k \hspace{-0.15cm}&=&\hspace{-0.15cm}-d u_k+\!(d-2+\eta_\phi)\rho_\phi u_k^{(1,0)} + \!(d-2+\eta_\chi)\rho_\chi u_k^{(0,1)}\nonumber\\
	 &&+I_{R,\phi}^d(\omega_\chi,\omega_\phi,\omega_{\phi\chi})+(N_1-1)I_{G,\phi}^d(u_k^{(1,0)})\nonumber\\
	 &&+I_{R,\chi}^d(\omega_\phi,\omega_\chi,\omega_{\phi\chi})+(N_2-1)I_{G,\chi}^d(u_k^{(0,1)})\,.\label{potflow}
\een
Herein, the first line arises from canonical dimensionality and the non-trivial wave-function renormalizations $Z_{\phi\, k}$ and $Z_{\chi\, k}$. The $u_k^{(1,0)}$ and $u_k^{(0,1)}$ denote the derivatives with respect to the first and second argument of $u_k$, respectively. The subsequent two lines correspond to the non-perturbative loop contributions of the massive radial and the Goldstone modes with factors $(N_1-1)$ and $(N_2-1)$. We have defined the threshold functions,
\ben
I_{R,i}^{d}(x,y,z)&=&\frac{4v_d}{d}\left(1-\frac{\eta_i}{d+2}\right)\frac{1+x}{(1+x)(1+y)-z},\nonumber\\
I_{G,i}^d(x)&=&\frac{4v_d}{d}\left(1-\frac{\eta_i}{d+2}\right)\frac{1}{(1+x)},
\een
with the volume element $v_d^{-1}=2^{d+1}\pi^{d/2}\Gamma(\frac{d}{2})$. The arguments in the flow equation \eqref{potflow} read
\ben
\omega_\phi&=&u_k^{(1,0)}+2\rho_\phi u_k^{(2,0)},\\
\omega_\chi&=&u_k^{(0,1)}+2\rho_\chi u_k^{(0,2)},\\
\omega_{\phi\chi}&=&4\rho_\phi \rho_\chi \big(u_k^{(1,1)}\big)^2.
\een
For the effective potential, we employ a polynomial expansion of the form
\begin{equation}\label{eq:expSSB}
u_k(\rho_{\phi}, \rho_{\chi})= \sum_{l+m\geq 2}^{n_{\mathrm{max}}} \frac{\lambda_{l,\, m}}{l! \cdot m!} \left(\rho_{\phi}- \kappa_{\phi} \right)^l\left(\rho_{\chi}- \kappa_{\chi}\right)^m,
\end{equation}
where $\kappa_{\phi}$ and $\kappa_{\chi}$ denote scale-dependent non-trivial vacuum expectation values in the symmetry-broken regime. For notational convenience, we do not indicate the scale-dependence of the $\kappa_\phi,\ \kappa_\chi$ and the $\lambda_{i,j}$ explicitely.

In order to derive explicit $\beta$ function for the couplings $\lambda_{i,j}$ and the expansion points $\kappa_\phi$ and $\kappa_\chi$, we have to specify the projection prescriptions. These can be straightforwardly derived from the scale-derivative of Eq.~\eqref{eq:expSSB} and read
\ben\label{eq:projections1}
\partial_t \kappa_\phi&=&\frac{u^{(0,2)}\partial_t u_k^{(1,0)}-u_k^{(1,1)}\partial_t u_k^{(0,1)}}{\left(u_k^{(1,1)}\right)^2-u_k^{(2,0)}u_k^{(0,2)}}\Big|_{\rho_\phi=\kappa_\phi \atop{\rho_\chi=\kappa_\chi}},\\
\partial_t \kappa_\chi&=&\frac{u_k^{(2,0)}\partial_t u_k^{(0,1)}-u_k^{(1,1)}\partial_t u_k^{(1,0)}}{\left(u_k^{(1,1)}\right)^2-u_k^{(2,0)}u_k^{(0,2)}}\Big|_{\rho_\phi=\kappa_\phi \atop{\rho_\chi=\kappa_\chi}},\\
\partial_t \lambda_{l,m}&=&\label{eq:projections2}\!\left(\partial_t u_k^{(l,m)}\!+\!u_k^{(l+1,m)}\partial_t\kappa_\phi\right.\!+\!\left. u_k^{(l,m+1)}\partial_t\kappa_\chi\right)\!\Big|_{\rho_\phi=\kappa_\phi \atop{\rho_\chi=\kappa_\chi}}\!.\nonumber\\ &&
\een
The flow equations for the wave-function renormalization factors $Z_{\phi}$ and $Z_{\chi}$ are derived by a suitable projection of the flow equation \eqref{eq:flowequwett} onto the momentum-dependent terms in the ansatz \eqref{eq:AnsatzEffectiveAction}, i.e.
\ben
\partial_t Z_{\phi}&=& (2\pi)^{d} \int\! d^{d}q \,\frac{\partial}{\partial p^2} \frac{\delta^{2}}{\delta \phi(p) \delta \phi(-q)} \partial_{t} \Gamma_{k} \vert_{\rho_\phi=\kappa_\phi \atop{\rho_\chi=\kappa_\chi}} , \\
\partial_t Z_{\chi}&=& (2\pi)^{d} \int\! d^{d}q\, \frac{\partial}{\partial p^2} \frac{\delta^{2}}{\delta \chi(p) \delta \chi(-q)} \partial_{t} \Gamma_{k} \vert_{\rho_\phi=\kappa_\phi \atop{\rho_\chi=\kappa_\chi}} .
\een
Note, that the functional derivatives with respect to the fields have to be specified. Typically, they will take into account both the contributions from the radial mode and massless Goldstone modes. Here, we do not distinguish between the two different contributions \cite{Berges:2000ew}.
In the following, we will restrict ourselves to a polynomial truncation to order $\phi^{8},\chi^{8}$ (LPA 8) or $\phi^{12},\chi^{12}$ (LPA 12), yielding a total of 14 or 27 couplings, respectively.

\begin{figure}[!here]
\includegraphics[width=0.95\linewidth]{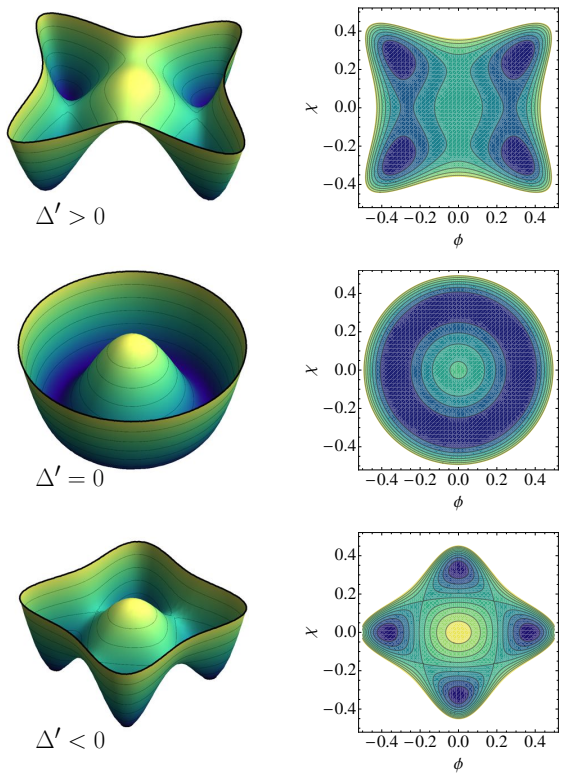}
\caption{\label{potplots}(Color online) Here we plot the potential $u(\phi, \chi)$ in 8th order LPA at a point with $\Delta' >0$ (upper two panels), $\Delta'=0$ (middle two panels) and $\Delta'<0$ (lower two panels). The second case exhibits an $O(2)$ rotation symmetry. For $\Delta'>0$ the minima lie at $\phi \neq 0$ and $\chi \neq0$, whereas they lie along one of the axes in field space in the case $\Delta'<0$.}
\end{figure}

At this point it is useful to re-examine the criterion $\Delta>0$ for our parameterization of the effective potential: It turns out that the position of the global minimum of the potential depends on the value of the quantity 
\be
\Delta' = \lambda_{2,0} \lambda_{0,2} - \lambda_{1, 1}^2,
\ee
which is clearly related to $\Delta$. The minimum lies at $\phi \neq 0, \chi \neq 0$ in the case $\Delta'>0$, and shifts onto one of the axes in field space, i.e., at $\phi =0, \chi \neq 0$ or $\phi \neq0, \chi=0$ for $\Delta'<0$, see Fig.~\ref{potplots}. The first case corresponds to a phase where both symmetries are broken. Accordingly this yields a tetracritical point, which is adjacent to this phase. In contrast, this phase does not exist for $\Delta'<0$, which implies that the multicritical point is bicritical in nature. 

The RG flow cannot cross the boundary $\Delta'=0$.  This follows immediately from symmetry considerations. Within theory space, the surface  $\Delta'=0$ defines a subspace with an enhanced symmetry (up to field re-parametrizations).
Such a symmetry-enhanced subspace is closed under RG transformation as long as the regulator-function respects the global symmetries. Accordingly any RG trajectory starting from a generic non-symmetric point can only approach the $\Delta'=0$ surface asymptotically, but can never cross it.

In the case $\Delta'<0$ it can be convenient to adapt the parametrization of the effective potential, Eq. \eqref{eq:expSSB}, to the physical situation when one of the expectation values vanishes. Then the parameterization \eqref{eq:expSSB} does not correspond to an expansion around the minima of the effective potential.
In this case we can employ the modified expansions
\begin{equation}
u_k(\rho_{\phi}, \rho_{\chi})=m_\chi^2 \rho_\chi+\sum_{l+m\geq 2}^{n_{\mathrm{max}}} \frac{\lambda_{l,\, m}}{l! \cdot m!} \left(\rho_{\phi}- \kappa_{\phi} \right)^l\rho_{\chi}^m,
\end{equation}
with the projection prescriptions
\ben
\partial_t m_\chi^2 &=& \left(\partial_t u_k^{(0,1)}+u_k^{(1,1)}\partial_t \kappa_\phi\right)
\Big|_{\rho_\phi=\kappa_\phi \atop{\rho_\chi=\kappa_\chi}}\,,\\
 \partial_t \kappa_\phi&=&-\frac{\partial_t u_k^{(1,0)}}{u_k^{(2,0)}}\Big|_{\rho_\phi=\kappa_\phi \atop{\rho_\chi=\kappa_\chi}}\,,
\een
 for the mass term $m_\chi^2$ and the minimum $\kappa_\phi$. For  the $\lambda_{i,j}$ the prescription in Eq. \eqref{eq:projections2} with $\partial_t \kappa_\chi=0$ holds.
%

\subsection{Relation to matrix models}\label{matrixmodels}

Here, we establish a correspondence between \mbox{$O(N_{1})\oplus O(N_{2})$}-models and certain matrix models (typically discussed in the context of condensed-matter theory\cite{Wegner:1979,Wegner:1979zz,Lee:1985zzc,Zirnbauer:1996zz,Roy:2011}, nuclear physics\cite{Brody:1981cx,Dyson:1962es}, QCD-like theories\cite{Osborn:1998qb,Shuryak:1992pi}, or two-dimensional quantum gravity \cite{Di Francesco:1993nw}; for a review see e.g.\ Ref.~\onlinecite{Guhr:1997ve}). A study of such models is possible with FRG tools\cite{Sfondrini:2010zm,Janssen:2012pq,mesterhazy2012,Jungnickel:1995fp,Berges:2000ew,Berges:1996ja,Bazzocchi:2011vr,Delamotte:2003dw,Fukushima:2010ji,Tissier:2000tz}.
Generally, matrix models can be phrased in terms of invariants of the reducible tensor representation of a given symmetry group. These invariants essentially describe the competing order parameters of the theory. Their identification relies on the decomposition of the tensor representation into irreducible representations that determine the possible symmetry breaking patterns of the symmetry group.

As an example consider the $U(2)$ matrix model written in terms of a Hermitian $2\times 2$ matrix, where the decomposition of the tensor representation $\overline{\mathbf{2}} \otimes \mathbf{2} = \mathbf{3} \oplus \mathbf{1}$ yields a coupled theory of two scalar fields with $SO(3) \oplus Z_{2}$ symmetry. The corresponding order parameters define the two invariants of the $U(2)$ matrix model that are written in terms of the trace in the defining representation $\bar{\sigma}_{1} = \tfrac{1}{2}\left(\textrm{Tr} \,\Phi\right)^{2}$ and $\bar{\sigma}_{2} = \tfrac{1}{2} \textrm{Tr} \,\Phi^{2}$. While the invariant $\sigma_{1}$ captures the breaking of the $Z_{2}$ center symmetry of the $O(3) \simeq SU(2)$ subgroup, a non-vanishing vacuum expectation value for the order parameter $\sigma_{2}$ leads to a breaking of the $SO(3)$ symmetry. The effective potential for the matrix model can be written solely in terms of these two invariants, i.e.,  $U(\Phi) = U(\bar{\sigma}_{1},\bar{\sigma}_{2})$. Higher order operators can be expressed completely in terms of linear combinations of $\bar{\sigma}_{1}$ and $\bar{\sigma}_{2}$ to some given power. Note that both invariants define field monomials of degree two and thus lead to the same type of competition for the corresponding order parameters, as discussed previously. If one examines the expansion of the potential in terms of these invariants
\begin{equation}
U(\Phi) = \sum_{l + m\geq 2} \frac{\bar{\lambda}_{l,m}}{l!\,m!} \left( \bar{\sigma}_{1} - \bar{\sigma}_{1,0} \right)^{l} \left( \bar{\sigma}_{2} - \bar{\sigma}_{2,0} \right)^{m} ~,
\end{equation}
where $\bar{\sigma}_{i,0}$ denote the corresponding expectation values in the symmetry broken phase, it is immediately apparent that this theory may exhibit a multicritical point that features an enhanced $O(4)$ symmetry. In fact, deriving the mass spectrum for this theory, one notices that it is completely equivalent to the coupled $SO(3) \oplus Z_{2}$ two-scalar model. This is an explicit example of universality -- the flow equations are completely independent of the field representations as long as the underlying symmetry and dimensionality of the problem are the same.

Let us use a different physical context to elucidate a subtlety in such matrix models:
Another prominent example featuring the $U(2)$ symmetry group appears in the context of low-energy effective models for QCD, e.g., the quark-meson model with two light quark flavors\cite{Jungnickel:1995fp}. It features a similar $SU(2)_{L} \times SU(2)_{R} \times U(1)_{A}$ symmetry which is written in terms of a generic complex matrix in the $\overline{\mathbf{2}} \otimes \mathbf{2}$ representation. Considering only the scalar sector, such a matrix theory can be written in terms of four invariants $\bar{\sigma}_{i} = \textrm{Tr}\, \left( \Phi^{\dagger} \Phi \right)^{i}$, $i = 1, \ldots, 4$, of the symmetry group. Similar to the discussion above, examining a polynomial expansion of the effective potential one could expect an enhanced $O(8)$ symmetry at the multicritical point. It turns out that here the competing order parameters do not necessarily enter with the same canonical mass dimension which may lead to different dynamics compared to the Hermitian $U(2)$ matrix model. In particular, there are no competing operators of degree two (the only mass-like invariant being $\textrm{Tr}\, \Phi^{\dagger}\Phi$) that are relevant in the critical domain. The situation is different however, in the case without the $U(1)_{A}$ axial symmetry, where an additional order parameter is allowed that violates this symmetry. It can be expressed in terms of a linear combination of $\textrm{det}\, \Phi$ and $\textrm{det}\, \Phi^{\dagger}$ and is obviously quadratic in the fields.

For the purpose of this paper it is useful to focus on $U(N)$ symmetric theories, where the corresponding tensor representation can be decomposed as \mbox{$\overline{\mathbf{N}} \otimes \mathbf{N} = (\mathbf{N^{2}-1}) \oplus \mathbf{1}$}. In that case, there are in general $N$ group invariants $\bar{\sigma}_{i}$, $i = 1, \ldots , N$ that define the order parameters and possible patterns of symmetry breaking. The number of group invariants depends on the rank of the group where the higher invariants essentially describe the possible breaking of the $O(N^{2}-1)$ symmetry. For this class of models we may exploit universality to map the flow equations onto the class of $Z_{2} \oplus O(N)$ symmetric theories. For that purpose one expands the potential only in terms of the respective (lowest) invariants (see Refs. \onlinecite{mesterhazy2012,Mesterhazy} for details). In this work, we will employ this correspondence to evaluate the anomalous dimension $\eta$ in specific cases, see sect.~\ref{IFP}.

\section{Results}\label{results}

\subsection{Fixed points from symmetry in $d=3$}

\subsubsection{Deducing fixed points from $O(N)$ models}

In this section, we deduce the existence and the properties of fixed points in three dimensions from symmetry arguments. These fixed points have been studied in great detail starting in Refs.~\onlinecite{KosterlitzNelsonFisher, NelsonKosterlitzFisher}.
We first observe that any subspace of theory space that shows an additional global symmetry must be a closed subspace under the RG flow, as long as we do not employ a symmetry-breaking regulator function. Accordingly any surface with an enhanced symmetry must be a fixed surface of the RG flow.
With the knowledge that models with $O(N)$ symmetry in three dimensions show a non-trivial, as well as a Gau\ss{}ian fixed point, one can immediately deduce the existence of 5 fixed points for the $O(N_1) \oplus O(N_2)$ model:
\begin{itemize}
\item A trivial, Gau\ss{}ian fixed point (GFP).
\item A decoupled fixed point (DFP), where the model decomposes into two disjoint $O(N_1)$ and $O(N_2)$ models and all mixed interactions such as $\lambda_{1,1}$ are zero. This fixed point must be tetracritical since $\Delta'>0$.
\item Two decoupled, semi-Gau\ss{}ian fixed points (DGFP), at which one of the $O(N_i)$ sectors approaches a Gau\ss{}ian, and the other a non-Gau\ss{}ian fixed point. Again, all mixed interactions vanish.
\item  A symmetry-enhanced isotropic fixed point (IFP), at which there is only one independent coupling at each order in the fields, e.g.,  $\lambda_{1,1}= \lambda_{2,0}= \lambda_{0,2}$, and the fixed-point coordinates agree with those of a $O(N_1+N_2)$ symmetric model.
\end{itemize}
In a condensed-matter setting, the fixed point with the lowest number of relevant directions, typically two, is commonly referred to as the stable one, as it has the least number of parameters that require tuning.
\begin{table}[!b]
\caption{FRG critical exponents for $O(N)$-models in three dimensions in a derivative expansion to order $\nabla^2$ and an expansion of the effective potential to order $\phi^{12}$ in comparison to the Monte Carlo results in Ref.~\onlinecite{HasenbuschA} for $N=1$, Ref.~\onlinecite{HasenbuschB} for $N=2$, Ref.~\onlinecite{HasenbuschC} for $N=3$ and  Ref.~\onlinecite{HasenbuschVicari}. These are FRG values obtained by the truncation and regularization scheme presented here and are employed to produce Fig.~\ref{n1n2plot}, \cite{MMStoappear}.}
\begin{ruledtabular}\label{ONtab} 
\begin{tabular}{ccccc}
N & $\nu$ &$\nu_{\rm MC}$  & $\eta$ & $\eta_{\rm MC}$\\ \hline
1 &  0.637 &0.63002(10) & 0.044 & 0.03627(10)\\
2& 0.685 & 0.6717(1) & 0.044 & 0.0381(2)\\
3 & 0.731 & 0.7112(5) & 0.041 &0.0375(5)\\
4&  0.772 & 0.750(2) & 0.037 & 0.0360(3)
\end{tabular}
\end{ruledtabular}
\end{table}

Dimensional analysis can help us to deduce the stability properties of the GFP and the two DGFPs: At the fully Gau\ss{}ian fixed point, 5 relevant parameters exist, it is therefore not likely to be the stable fixed point. At a (partially) interacting fixed point, fluctuations contribute to the scaling dimensions of operators, which accordingly depart from canonical scaling, and can even change their sign. In contrast, at a fully Gau{\ss}ian fixed points, critical exponents are determined by canonical scaling. The two DGFPs show an interesting property when it comes to the question of relevant couplings, which we will discuss further in subsect.~\ref{DGFPdiscussion}. For a vanishing coupling, one might at a first glance expect its critical exponent to agree with its dimensionality. It turns out that the critical exponents of two-scalar interaction couplings such as $\lambda_{1,1}$ receive a non-trivial contribution from fluctuations and do not accord with a naive dimensional analysis. Thus dimensional analysis only implies the existence of at least two relevant couplings at this fixed point. Using that a $O(N)$ model has one relevant coupling, see tab.~\ref{ONtab}, we conclude that the DGFPs have at least three relevant directions and are therefore not likely to be the stable ones.
For the IFP and the DFP, we can infer a subset of their critical exponents from $O(N)$ models, and conclude that the IFP has at least one, and the DFP at least two relevant couplings. To determine which of these is the stable one, a detailed analysis of their critical exponents is necessary. In the following, we will conduct a numerical search for fixed points and determine their critical exponents numerically.
%

\subsubsection{Critical exponents for the symmetry-enhanced IFP}\label{IFP}

At the IFP, a subset of the critical exponents can be inferred directly from those of an $O(N)$ symmetric theory where $N=N_1+N_2$. Besides the appertaining eigendirections, which correspond to an $O(N)$  symmetric approach to the fixed point, further exponents exist that cannot be inferred from $O(N)$ models. In other words, the theory space of an $O(N)$ model corresponds to a genuine subspace of the $O(N_1) \oplus O(N_2)$ model. Therefore the universality class of the symmetry-enhanced $O(N)$ fixed point is a particular enlargement of the well-known  $O(N)$ universality class. 

Interestingly, a further universality class exists, corresponding to a theory space with an additional $\phi \leftrightarrow \chi$ symmetry, existing for $N_1=N_2$, which contains the $O(N)$ theory space while itself being embedded in the full $O(N_1) \oplus O(N_2)$ theory space. Our theory space thus corresponds to a nesting of three closed theory spaces: The smallest $O(N)$ theory space lies within the $\phi\leftrightarrow \chi$ symmetric theory space, which itself is embedded in the full $O(N_1) \oplus O(N_2)$ theory space. The same nesting pattern holds for the critical exponents: A subset of the critical exponents agrees with all critical exponents of the $\phi\leftrightarrow\chi$ symmetric case. A subset of these then agrees with all critical exponents of the $O(N)$ model. It turns out that the $\phi \leftrightarrow \chi$ symmetric universality class is of particular interest when it comes to determining the IR stability of the fixed point: Inheriting one positive critical exponent from the $O(N)$ model for all values of $N$, cf. $\theta_2$ in tab.~\ref{ONtab}, and showing a second positive one for all $N$, cf. $\theta_1$ in tab.~\ref{ONtab}, it is the third-largest critical exponent of the universality class which determines how many parameters need to be tuned in order to reach a second-order phase transition. This particular critical exponent changes its sign from negative to positive between $N=2$ and $N=3$, and thus changes the stability properties of this fixed point. To determine its value, it actually suffices to consider the $\phi \leftrightarrow \chi$ symmetric model. Dropping this symmetry-restriction only adds further critical exponents to the universality class, cf. 
$\theta_1$ and $\theta_4$ in 
Tab.~\ref{Nctab}, but does not change the values of the other exponents. 
\begin{table}[!here]
\caption{Table with critical exponents of the IFP in LPA 12 including the anomalous dimension $\eta$. For comparison, we also show the $y_{i,j}$-notation from Ref.~\onlinecite{Calabrese:2002bm}. The crossover exponent is given by $\phi_T=y_{2,2}\nu$ where $\nu=1/\theta_2$ is the exponent of the correlation length. The $O(N)$ critical exponents are highlighted in italics.}
\begin{ruledtabular}\label{Nctab}
\begin{tabular}{cccccc|c}
$N$& $\theta_1=y_{2,2}$& $\theta_2=y_{2,0}$& $\theta_3=y_{4,4}$& $\theta_4=y_{4,2}$& $\theta_5=y_{4,0}$ & $\phi_T$ \\ \hline
2 & 1.756 & {\it 1.453} & -0.042 & -0.446 & {\it -0.743} & 1.209 \\
2.31 & 1.767 & {\it 1.423} & -0.0009 & -0.425 & {\it -0.746} & 1.242 \\
3 & 1.790 & {\it 1.362} & 0.086 & -0.380 & {\it -0.756} & 1.314 \\
4 & 1.818 & {\it 1.292} & 0.196 & -0.324 & {\it -0.775} & 1.407 \\
5 & 1.842 & {\it 1.240} & 0.289 & -0.283 & {\it -0.797} & 1.485 \\
10 & 1.908 & {\it 1.116} & 0.568 & -0.154 & {\it -0.879} & 1.710 \\
100 & 1.990 & {\it 1.010} & 0.951 & -0.015 & {\it -0.988} & 1.970\\
1000 & 1.999 & {\it 1.001} & 0.995 & -0.002 & {\it -0.999} & 1.988\\
\hline
$\infty$ & 2 & {\it 1} & 1 & 0 & {\it -1} & 2
\end{tabular}
\end{ruledtabular}
\end{table}

The value $N_c$, at which $N$ the third-largest critical exponent of the IFP changes its sign, has been a much-investigated question, cf. Refs.~\onlinecite{Calabrese:2002bm, Moser2008} and references therein. Here, we present a first analysis using the nonperturbative functional RG. Our results indicate, in accordance with Refs.~\onlinecite{Calabrese:2002bm, Moser2008} that the IFP has two positive critical exponents for $N=2$, and three for $N=3$. We find that the transition occurs for a critical value of $N_c \approx 2.32$, see tab.~\ref{Nctab}. This should be compared to a value of $N_c =2.89(4)$ from a six-loop calculation \cite{Carmona2000} and $N_c\approx 2.6$ in Ref.~\onlinecite{Moser2008}.
Let us compare our results to those obtained in  Ref. \onlinecite{HasenbuschVicari} using Monte Carlo simulations. There, $y_{2,2}=1.7639(11)$ for $N=2$, $y_{2,2}=1.7906(3)$ for $N=3$ and $y_{2,2}=1.8145(5)$ for $N=4$. These compare rather well to our results. For the case of $y_{4,4}$, the Monte-Carlo values are $y_{4,4}=-0.108(6)$ for $N=2$, $y_{4,4}=0.013(4)$ for $N=3$ and  $y_{4,4}=0.125(5)$ for $N=4$. Here, our values are larger, and we expect to obtain better precision at higher orders in the truncation. 

Further, we explicitely compare our results for the crossover exponent $\phi_T=\theta_1/\theta_2=y_{2,2}\nu$ in the case $N=3$ to the ones obtained by other approaches: Here, we get $\phi_T(N=3)=1.314$ which is in reasonable agreement with $\phi_T=1.260$ from the five-loop $\epsilon$-expansion in Ref.~\onlinecite{Calabrese:2002bm} or $\phi_T=1.275$ from the two-loop RG series in Ref.~\onlinecite{Moser2008}. We obtain a similar comparison for other choices of $N$. While the largest critical exponents $\theta_1,\theta_2$ already show good quantitative precision, the values for the subleading exponents $\theta_3,\theta_4,\theta_5$ are less accurate. Precision is expected to be achieved by an extension of the truncation towards higher orders in the derivative expansion as, e.g., shown for scalar models in Refs.~\onlinecite{Canet2003,Litim:2010tt}: For instance, the seven-loop result for the critical exponent of the correlation length, $\nu=0.6304(13)$, cf. Ref.~\onlinecite{zinnjustin1989} compares well to the FRG estimate in fourth order of the derivative expansion $\nu=0.632$, cf. Ref.~\onlinecite{Canet2003}.

To study the effect of terms beyond our truncation, it is useful to vary the numerical value for $\eta$ by hand and test the variation of $N_c$. Although this is not a self-consistent solution to the set of flow-equations, it can provide a measure for the sensitivity of results on terms beyond the truncation. Assuming that $\partial_{\lambda_i} \eta(\lambda_i)|_{\lambda_i=\lambda_{i\, \ast}}\ll1$, we obtain a variation of $N_c$ of $\sim 0.25$ when varying $\eta \in (0, 0.1)$.

In a previous application of the FRG method to a related $N$-vector model with cubic anisotropy, the critical value for $N_c$ has been determined to be $N_c=3.1$ \cite{Delamotte2002}, working with an exponential instead of an optimized regulator shape function. This study works with a truncation up to order eight in the fields and including derivative terms up to second order in momenta and fourth order in the fields. In our truncation, we actually observe a change of $~0.1$ in the value of $N_c$, when going from LPA 8 to LPA 12. An analysis of the same model with perturbative tools yields $N_c<3$, cf. Refs.~\onlinecite{Shalaev, Kleinert, Mayer, Carmona2000, Folk2,Varnashev1, Varnashev2, Folk,Newman}.

\subsubsection{Critical exponents for the DFP}\label{DFP}

The stability analysis of the DFP simplifies due to an exact scaling relation \cite{Aharony2, Grinstein, Aharony3}. Four of the critical exponents correspond to the $O(N_1)$ and $O(N_2)$ critical exponents $\theta_1 = \frac{1}{\nu_1}$, $\theta_2 = \frac{1}{\nu_2}$, $\theta_4 =- \omega_1 $ and $\theta_5= - \omega_2$. The third, which actually decides about the stability, is given by the relation
\begin{equation}\label{eq:scarel}
\theta_3=\frac{1}{\nu_1}+\frac{1}{\nu_2}-d\,.
\end{equation}
This allows us to determine the critical exponents for the DFP from the pure $O(N)$ model for which the expression for the anomalous dimension $\eta$ is known, see Ref.~\onlinecite{Berges:2000ew} for details. Using calculations in LPA to 12th order including a simple approximation to $\eta$, determined with respect to the Goldstone modes, we get the following table for the critical exponents, see Tab.~\ref{DFPcritexps}. 
\begin{table}[!here]
\caption{We show the five largest critical exponents of the DFP as a function of $N_1$ and $N_2$ in 12th order LPA including the anomalous dimension.}
\begin{ruledtabular}\label{DFPcritexps} 
\begin{tabular}{cccccccc}
$N$&$N_1$& $N_2$ & $\theta_1$ & $\theta_2$ & $\theta_3$ & $\theta_4$ & $\theta_5$\\
\hline
2& 1 & 1 & 1.571 & 1.571 &0.142& -0.728& -0.728 \\
3& 1& 2& 1.571 & 1.459 & 0.030& -0.728 & -0.735\\
4& 1& 3& 1.571 & 1.367 & -0.062& -0.728 & -0.748 \\
4& 2 &2& 1.459 & 1.459 & -0.082& -0.735 & -0.735 \\
5& 1& 4& 1.571& 1.296 &-0.133& -0.728 & -0.768 \\
5& 2&3& 1.459 & 1.367  & -0.174& -0.735 & -0.748
\label{DFPcritexp}
\end{tabular}
\end{ruledtabular}
\end{table}

Our results are in accordance with Refs.~\onlinecite{Calabrese:2002bm,Moser2008}. For $N_1 =1$ we obtain a critical value of $N_2=2.31$, to be compared, e.g., to $2.17$ from Ref.~\onlinecite{Moser2008}.
We conclude that the case $N_1=2, N_2=3$ might be relevant for high-${\rm T_c}$ superconductors, features a stable tetracritical DFP, see Ref.~\onlinecite{Zhang1997, Zhang2}.

To test the quality of our truncation, we can compare the value for $\theta_3$ as obtained from the scaling relation to the result from the explicit diagonalization of the stability matrix. Within a truncated RG flow, we do not expect the exact scaling relation to be fulfilled precisely. Here, we observe that a determination of $\theta_3$ by explicit diagonalization shifts the transition line for the stability of the DFP in the $N_1-N_2$-plane by an absolute value $\sim 0.01$ within a fixed order of the truncation.

\subsubsection{Stable fixed points as a function of $N_1$ and $N_2$}

\begin{figure}[!here]
\includegraphics[width=0.7\linewidth]{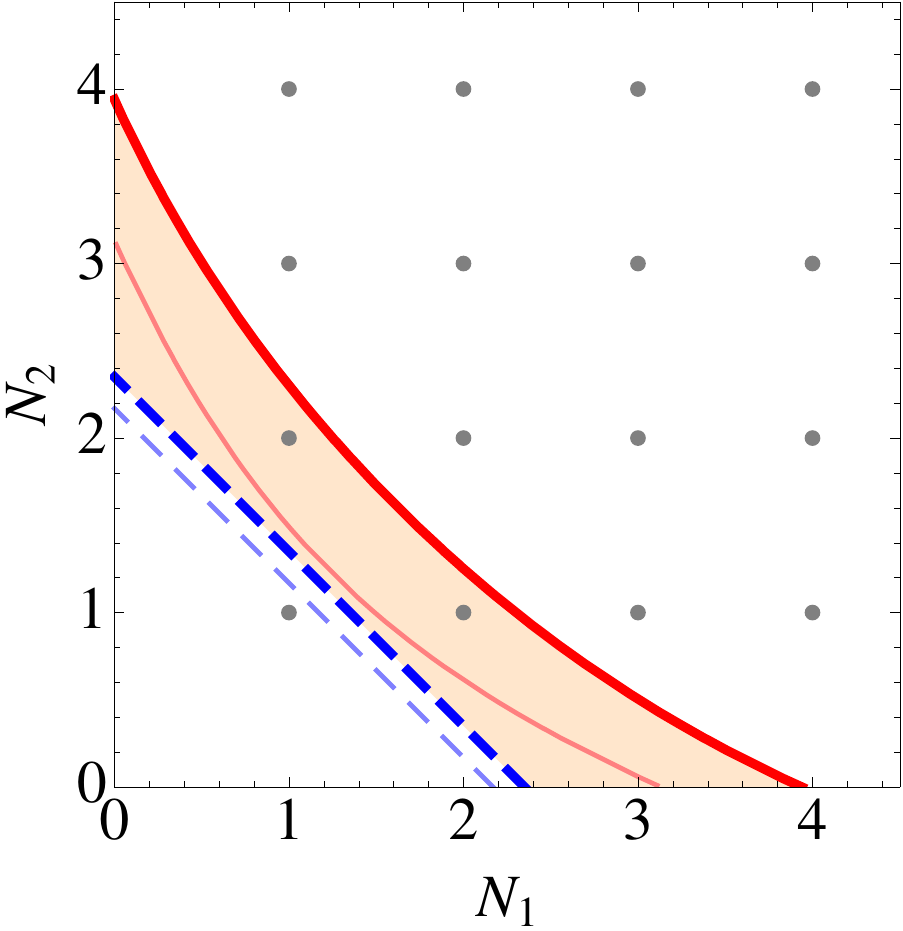}
\caption{\label{n1n2plot}(Color online) For LPA 12 including $\eta \neq0$, the IFP is stable for $N_1+N_2 <2.32$ (thick, blue dashed line), and the DFP is stable for values to the upper right of the red full line. The thin lines denote the stability boundaries in LPA 8 with $\eta=0$.}
\end{figure}

Here, we refer to a fixed point as being stable when it shows exactly two relevant directions. From the two previous Subsects.~\ref{IFP} and \ref{DFP}, it is possible to deduce that the IFP is stable up to $N_c = N_1+N_2 \approx 2.32$, and the DFP is stable for $N_1, N_2$ beyond a critical line, which depends on $N_1$ and $N_2$ separately, see Fig.~\ref{n1n2plot}.
This result is in good agreement with Ref.~\onlinecite{Moser2008}. The shaded region in Fig.~\ref{n1n2plot} contains the physical points $N_1=1, N_2 =2$ and $N_1=2, N_2=1$, where neither of the fixed points deduced from symmetry is stable. This leads us to investigate whether additional fixed points that do not follow from the $O(N)$ Wilson-Fisher fixed point can exist in these models.
%

\subsection{Additional fixed points in $d=3$}

\subsubsection{Biconical fixed point}

Apart from those fixed points that can be inferred from $O(N)$ models, the interaction between the two sectors could induce further non-trivial fixed points. Indeed, an additional fixed point, termed the biconical one (BFP), has first been discovered in Ref. \onlinecite{NelsonKosterlitzFisher} and further studied in Refs.~\onlinecite{Calabrese:2002bm, Moser2008}.

\begin{figure}[!here]
\includegraphics[width=\linewidth]{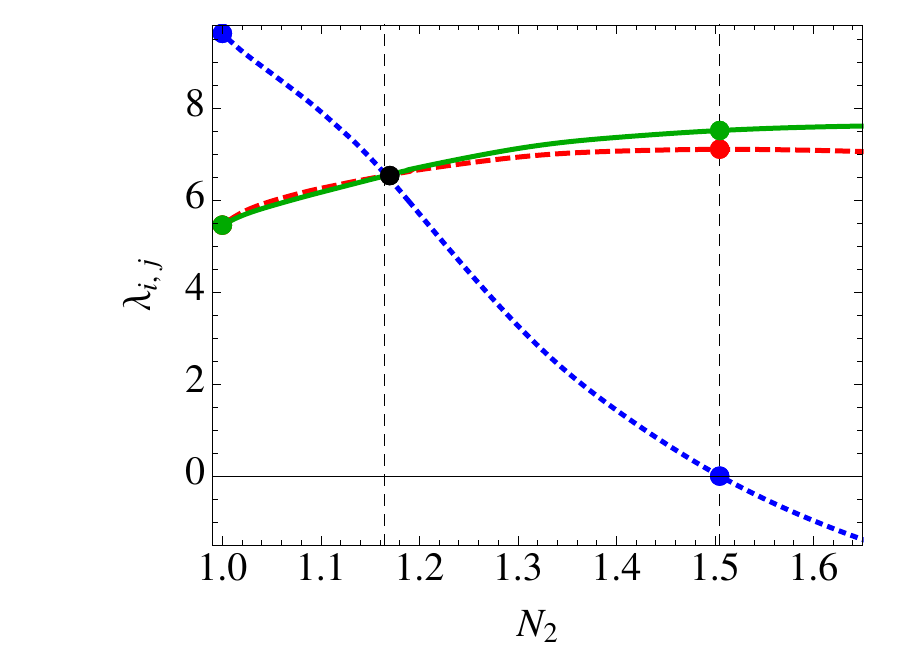}
\includegraphics[width=\linewidth]{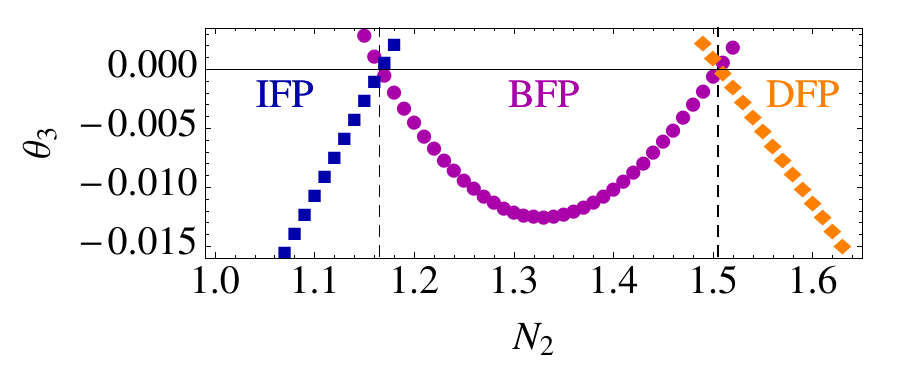}
\caption{\label{BFPplot}(Color online)  Here we plot the fixed-point values (upper panel) of $\lambda_{2,0}$ (green solid line), $\lambda_{0,2}$ (red dashed line) and $\lambda_{1,1}$ (blue dotted line) as a function of $N_2$ for $N_1=1$ at the BFP. The third largest critical exponent, deciding about the stability of the fixed point, is negative for the BFP between $1.17 <N_2< 1.5$ (purple discs in the lower panel). For $N_2 \lesssim 1.17$, the IFP is stable (blue squares), and for $N_2 >1.5$, the IFP becomes stable (orange diamonds). Note that $\lambda_{1,1}<0$ does not automatically yield an unstable potential, as long as this coupling does not exceed a critical value.}
\end{figure}

Here, we search for this fixed point in 8th order LPA without anomalous dimensions, i.e., $\eta_{\phi} = \eta_{\chi}=0$. As is clear from the thin lines in Fig.~\ref{n1n2plot}, this fixed point should exist and be stable in the region $1.17 \lesssim N_2 \lesssim 1.50$ for $N_1=1$, at this order of the approximation. The BFP emerges from the IFP (see upper panel of Fig.~\ref{BFPplot}) and immediately becomes the stable fixed point, see lower panel in Fig.~\ref{BFPplot}. At $N_2 \approx 1.5$, it merges with the DFP, which, in this order of the approximation, is the stable fixed point beyond this value. Note that the critical exponent $\theta_3$ for the DFP in Fig.~\ref{BFPplot} has been obtained by direct diagonalization of the stability matrix and not by the scaling relation, Eq.~\eqref{eq:scarel}, to be consistent with the determinantion of $\theta_3$ for the BFP and the IFP. We observe that $\Delta'>0$ in the stability region of the BFP. This implies tetracritical behavior in that region.

As is obvious from Fig.~\ref{n1n2plot}, the stability region of the DFP as found by resorting to the pure $O(N)$ model with anomalous dimensions begins for larger values of $N_1, N_2$ than found from LPA 8 without anomalous dimensions. We expect that, if we extend our results for the BFP to non-vanishing $\eta$, the region of existence and stability of the BFP becomes wider. In agreement with the conjecture that there is always one stable fixed point  \cite{KosterlitzNelsonFisher, NelsonKosterlitzFisher, Brezin}, we expect the BFP to be stable in the complete shaded region in Fig.~\ref{n1n2plot}.
Thus the physical point $N_1=1,\ N_2 =2$ should indeed be described by the BFP as its stable fixed point. As we have shown using LPA 12 with $\eta\neq0$, neither the IFP nor the DFP are stable at this point.
We defer a more detailed study of this physically interesting situation including explicit expressions for the anomalous dimensions of the two bosonic sectors $\eta_\phi$ and $\eta_\chi$ to future work. 

\subsubsection{Fixed points for $\Delta'<0$}

We observe that the BFP moves into the region $\Delta'<0$ after colliding with the IFP,  as $\lambda_{1,1}> \lambda_{2,0/0,2}$, cf. Fig.~\ref{BFPplot}. In particular, it approaches another symmetry-enhanced fixed point for $N_1=1=N_2$. This motivates us to analyze the existence of fixed points for $\Delta'<0$ in more detail. 

Let us specialize to the case $N_1=N=N_2$ to discuss the possible existence of further fixed points:
Here, we observe an additional fixed point with an enhanced discrete symmetry $\phi \leftrightarrow \chi$, as has also been discussed as the cubic fixed point in Refs.~\onlinecite{Bornholdt:1994rf,Bornholdt:1995rn, Delamotte2002}. We call it the symmetric fixed point (SFP). The existence of this fixed point can be inferred as follows: As discussed above, the case $\Delta'<0$ corresponds to a situation where the four degenerate minima of the potential lie along  the axis $\phi=0$ or $\chi=0$. Rotating the basis $(\phi, \chi)$ in field space to a new basis $(\tilde{\phi}, \tilde{\chi})$, which is tilted by $\pi/4$ yields a potential that is symmetric in $\tilde{\phi}, \tilde{\chi}$ and has minima that lie along the diagonal, see Refs.~\onlinecite{Bornholdt:1994rf,Bornholdt:1995rn}. In fact it is possible, by a redefinition of the couplings, to recover precisely the form
\eqref{eq:expSSB}. Accordingly, we can again deduce the existence of fixed points from the existence of $O(N)$ universality classes. 
This implies the existence of a new fixed point, which is a decoupled fixed point in the $\tilde{\phi}, \tilde{\chi}$ coordinates and features a $\phi \leftrightarrow \chi$ symmetry in the original basis. As it corresponds to the DFP in the new coordinates, with the additional $\phi \leftrightarrow \chi$ symmetry imposed on the couplings, we infer that three of the critical exponents are identical to this case, see Ref.~\onlinecite{Bornholdt:1995rn}. The additional two critical exponents that arise from relaxing the symmetry $\phi \leftrightarrow \chi$ cannot be deduced from the DFP. Consequently, this fixed point has at least three relevant directions for $N=1$. (At least) one of the critical exponents then changes sign for $N=2$, see tab.~\ref{DFPcritexp}. We therefore conclude, that a potentially stable fixed point exists at $\Delta'<0$, implying bicritical behavior.

Besides, we find a fixed point with non-trivial interaction between the two sectors, and no enhanced symmetry, which we call the asymmetric fixed point (AFP), see tab.~\ref{AFPtab}. We find this fixed point in an expansion of the effective potential around one of the saddle points  which is a consequence of our ansatz \eqref{eq:expSSB}. We do not expect such an expansion to yield accurate critical exponents. An improved estimate for their values can be obtained by implementing the full effective potential with a higher numerical effort, which is beyond the scope of the present work.
\begin{table}[!here]
\caption{Fixed-point values and real parts of the critical exponents for the asymmetric fixed point as a function of the expansion order $n_\mathrm{max}$ of the effective potential. At the highest order of the LPA, all five critical exponents become real.\label{AFPtab}}
\begin{ruledtabular}\begin{tabular}{cccccc}
$n_\mathrm{max} $ & $\kappa_\phi$ & $\kappa_\chi$ & $\lambda_{2,0} $ & $\lambda_{0,2} $ & $\lambda_{1,1} $ \\ \hline
4 &  0.00871 & 0.0459  &  10.662 & 0.384  &  6.069\\
8 & 0.0135  &  0.0172 & 5.543  &  3.315 & 10.573 \\
12 & 0.0146  &  0.0167 & 5.162  &  4.000 & 10.383 \\ \hline
$n_\mathrm{max} $ & $\theta_1$ & $\theta_2$ & $\theta_3$ & $\theta_4$ & $\theta_5$ \\ \hline
4 & 1.800  & 1.220  &  1.009 &  -0.900 &  -0.900\\
8 & 2.037  & 1.556  &  0.148 &  -0.267 &  -0.267 \\
12 & 1.990  & 1.523  &  0.150 & -0.042 &  -0.732 
\end{tabular}\end{ruledtabular}
\end{table}

Note that this fixed point has three relevant parameters. Accordingly, it simply may not be possible to reach it in a realistic experimental setting, since three tunable parameters might not be available. In this case, one might observe nonuniversal behavior where the system undergoes a first order phase transition. A similar situation applies for the SFP. In that case systems with initial conditions in the $\Delta'<0$ region would still show first-order phase transitions, even though fixed points in that region exist.
Let us clarify this statement in some more detail: For a second-order phase transition, the existence of a fixed point is a necessary, but not a sufficient condition. It is necessary, as the divergence of the correlation length at a second-order phase transition corresponds to scale-freedom in the system and can therefore only be realized at a fixed point. If no fixed point exists, no second-order phase transition can occur.
For a given physical system, the existence of a fixed point is not sufficient for the second-order phase transition to occur: To reach a fixed point, the couplings corresponding to relevant operators need to be tuned. These correspond to physical parameters of the system that would require tuning in order to reach the second-order phase transition. In a given system, some of these parameters might actually not be tunable (e.g., if they correspond to fixed microscopic parameters of the system). In that case, the fixed point cannot be reached, and the possibility of the second-order phase transition is never realized. The more relevant couplings exist, the more likely this situation is. We thus conclude that the existence of the AFP and the SFP need not necessarily imply that the corresponding physical systems undergo second-order phase transitions.

Let us discuss the reason why the AFP has not been discovered by perturbative tools: Expanding our $\beta$ functions for small values of the interaction couplings and around a vanishing vacuum-expectation value, we get a perturbative approximation to our full system. It turns out that the AFP is not a fixed point of these perturbative $\beta$ functions. We conclude that it is nonperturbative in nature, and probably connected to threshold effects in the $\beta$ functions which are invisible to perturbation theory. In contrast, a nonperturbative Monte Carlo study should be able to access the AFP and confirm or refute its existence.
Employing the FRG beyond a polynomial expansion of the potential will also yield a non-trivial test of the existence of these fixed points. At present, our study cannot preclude the possiblity that this fixed point arises as a truncation artefact.

\begin{figure}[!here]
\includegraphics[width=1.0\linewidth]{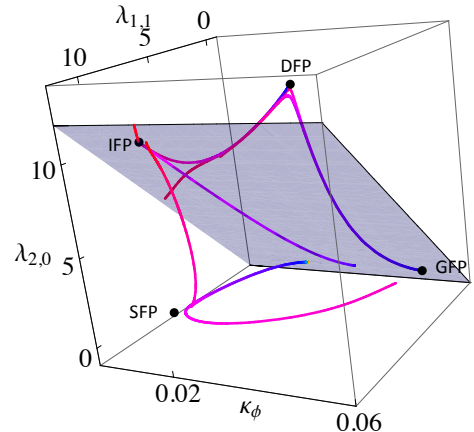}
\caption{\label{trajectoriesplot}(Color online)  Here we show several selected trajectories in the symmetry-reduced setting with $\phi \leftrightarrow \chi$ in LPA 4, which connect the fixed points. Due to the symmetry requirement, the AFP does not appear. The trajectories start in the UV (purple color in online version) and flow toward the IR (red color in online version). None of the shown trajectories cross the $\Delta' = 0$ plane.}
\end{figure}

The SFP and the AFP are both characterized by $\Delta'<0$, thus corresponding to bicritical rather than tetracritical behavior according to the mean-field criterion. Thus, contrary to the analysis in Refs.~\onlinecite{Calabrese:2002bm, Moser2008}, the RG flow within our truncation shows fixed points in both the $\Delta'>0$ as well as the $\Delta'<0$ region. Since the RG flow does not cross the $\Delta'=0$ boundary (see Fig.~\ref{trajectoriesplot}), the non-existence of fixed points for $\Delta'<0$ would imply that initial conditions within this region must necessarily lead to a first-order phase transition. Our results suggest that in fact second-order phase transitions could also exist in this region. Of course it remains to confirm the existence of these fixed points beyond our truncation. 


To summarize, the RG flow features a total of seven fixed points. For the $\phi \leftrightarrow \chi$ symmetric case, these reduce to four fixed points, which are shown in Fig.~\ref{trajectoriesplot} for the case $N_1=1=N_2$.

\subsection{Relevance for high-energy physics}

Several features arise in the two-scalar $Z_2 \oplus Z_2$ model with potential interest for high-energy model, which we will discuss in the following.

\subsubsection{Critical exponents for the semi-Gau\ss{}ian fixed point}\label{DGFPdiscussion}

At the DGFPs only a small subset of couplings is non-zero, namely only the self-interaction of one scalar. The self-interactions of the second scalar, as well as the interactions between the two sectors vanish.
Nevertheless, the corresponding critical exponents are non-trivial and do not all follow from canonical dimensionalities.
It turns out that non-trivial scaling arises due to the interactions between the two sectors, even if, at the fixed point, $\lambda_{1,1}=0$, and similarly for higher-order couplings. Diagrammatically, the $\beta$ functions of these couplings receive contributions from diagrams where (some of the) vertices are proportional to the couplings in the interacting scalar sector. These yield non-canonical entries in the stability matrix.
This clearly suffices to give non-trivial critical exponents, see Tab.~\ref{DGFPtab}.
\begin{table}[!b]
\caption{\label{DGFPtab} Here we show the real part of the five largest critical exponents at the DGFP at order $n_\mathrm{max}$ in LPA. The italic values are those of the Ising universality class.}
\begin{ruledtabular}\begin{tabular}{cccccccccc}
$n_\mathrm{max}$& $\theta_1$ & $\theta_2$ & $\theta_3$ & $\theta_4$ & $\theta_5$ &$\theta_6$ &$\theta_7$ & $\theta_8$& $\theta_9$\\
\hline
4& \it{2}& 1&1&0.667&\it{ -0.333}& - & - & - & -\\
6& \it{1.372} &1&1&0.547&0&-0.291 &-0.928&\it{-1.075} &-1.075.
\end{tabular}
\end{ruledtabular}
\end{table}
Reaching the Ising model as a special decoupled point in a larger theory space that features interactions between the two sectors therefore yields a new, non-trivial universality class, which is not simply constructed from the Ising universality class and canonical critical exponents.

We accordingly observe a rather interesting property of an interacting fixed point: Even if a sector of the theory is fully non-interacting at this fixed point, this does not necessarily imply canonical critical exponents.
Let us for a moment indulge in speculation, and assume a scenario, in which the standard model is UV complete with the help of an interacting fixed point, known as the asymptotic-safety scenario \cite{Weinberg:1980gg}, e.g., induced by the coupling to an asymptotically safe quantum theory of gravity \cite{Reuter:1996cp,Percacci:2002ie,Percacci:2003jz,Daum:2009dn,Daum:2010bc,Folkerts:2011jz,
Harst:2011zx,Zanusso:2009bs,Vacca:2010mj,Shaposhnikov:2009pv,Narain:2009fy,
Narain:2009gb,Eichhorn:2011pc,Eichhorn:2012va}. In this case, even asymptotically free sectors of the theory could show non-trivial scaling behavior. In analogy to the DGFP in our model, where the interacting sector induces nontrivial scaling in the noninteracting sector, gravitational fluctuations could yield noncanonical scaling for asymptotically free matter couplings.

\subsubsection{Emergence of Goldstone modes from discrete symmetries}

\begin{figure}[!here]
\includegraphics[width=0.9\linewidth]{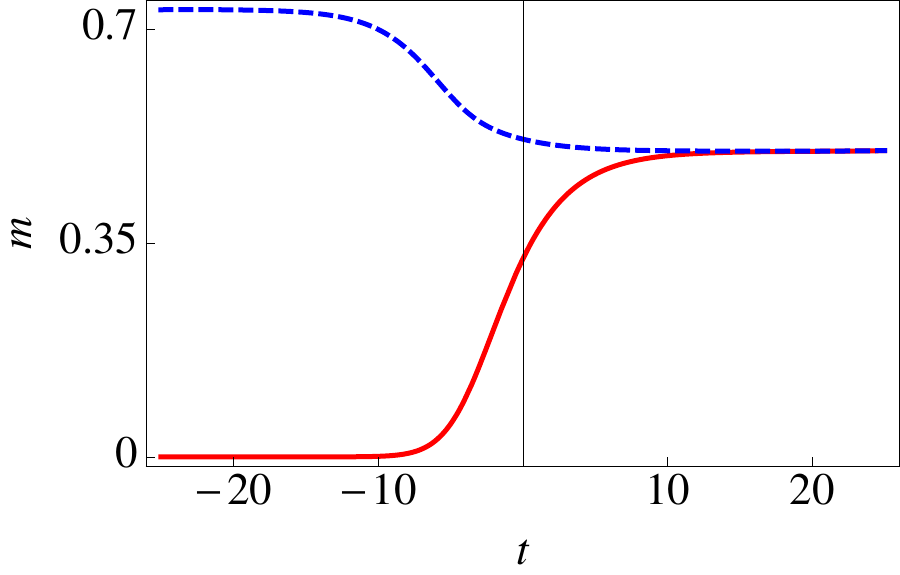}
\caption{\label{massflow}(Color online)  Here, we plot the two mass eigenvalues of the $u^{(2)}$ matrix, corresponding to the dimensionless masses $m_i(k) = \frac{\bar{m}_i}{k}$ of physical excitations as a function of the RG scale $t= {\rm ln} (k/\Lambda)$, on a trajectory connecting the DFP in the UV (high $t$) with the IFP in the IR (low $t$).}
\end{figure}

The emergence of massless modes from symmetry breaking is expected from continuous symmetries, only. There is no Nambu-Goldstone mode associated with the breaking of a discrete symmetry. Accordingly, one would naively expect that the $Z_2 \oplus Z_2$ model does not exhibit massless modes. This is actually not the case, see Fig.~\ref{massflow}. There we show the two masses as a function of the RG scale on a particular RG trajectory connecting the DFP in the UV with the IFP in the IR in $d=3$. At low momenta, one of the masses clearly goes to zero, thus a massless mode emerges.
This is in fact in full accordance with Goldstone's theorem, since there is one particular point in our theory space which is characterized by a continuous symmetry, namely $O(2)$. It contains a fixed point of the RG flow, namely the IFP. Accordingly there are RG trajectories which approach the IFP asympotically in the IR. The previous analysis of the stability properties has shown that for the $Z_2 \oplus Z_2$ model the IFP is the stable fixed point. Starting the RG flow in the symmetry-broken regime, Goldstone's theorem then forces one of the two massive modes to become massless in this limit. As the corresponding fixed point at which the symmetry-enhancement takes place shows IR-repulsive directions, finetuning is required in order to reach this fixed point and observe the emergence of Goldstone modes.

Let us add that in principle a similar mechanism might be invoked to generate a mass hierarchy in a system where the microscopic Lagrangian contains equal masses. If the system can exhibit an enhancement of the symmetry by an additional continuous symmetry transformation, then a spontaneous breaking of this additional symmetry in the infrared must produce a massless Goldstone mode. Small explicit symmetry breaking terms can then give a small mass to this pseudo-Goldstone mode. Compared to the other masses in the theory the pseudo-Goldstone boson mass could then remain rather small, thus producing a hierarchy.

\subsubsection{Transition to $d=4$ dimensions}

The $O(N)$ theory suffers from the triviality problem in $d=4$ dimensions: The theory shows no interacting fixed point and the Gau\ss{}ian fixed point is IR stable. Towards the UV, the scalar self-coupling diverges at a finite scale, i.e., it shows a Landau pole. This happens, unless the interaction is tuned to vanish in the IR, yielding a trivial theory. Whether the Higgs sector of the Standard Model shows this problem is not fully clear, as a growing Higgs self-coupling implies that the theory enters a strongly-interacting regime, where further fermion or gauge boson fluctuations might potentially prevent the Landau pole, see, e.g., Refs.~\onlinecite{Gies:2009hq,Gies:2009sv,Bazzocchi:2011vr}. A second possibility could be given by an interacting theory of two scalars, where fluctuations of the second scalar field could counteract those of the first scalar, and thus suppress the Landau pole. One usually expects that this option is not realized in four dimensions. Here, we provide further evidence for the triviality of coupled two-scalar models, by following the fate of the fixed points towards $d=4$. Our result implies that, e.g., Higgs-inflaton theories as well as hybrid models for inflation with two coupled scalar fields \cite{Linde:1993cn} both suffer from the triviality problem.

\begin{figure}[!here]
\includegraphics[width=\linewidth]{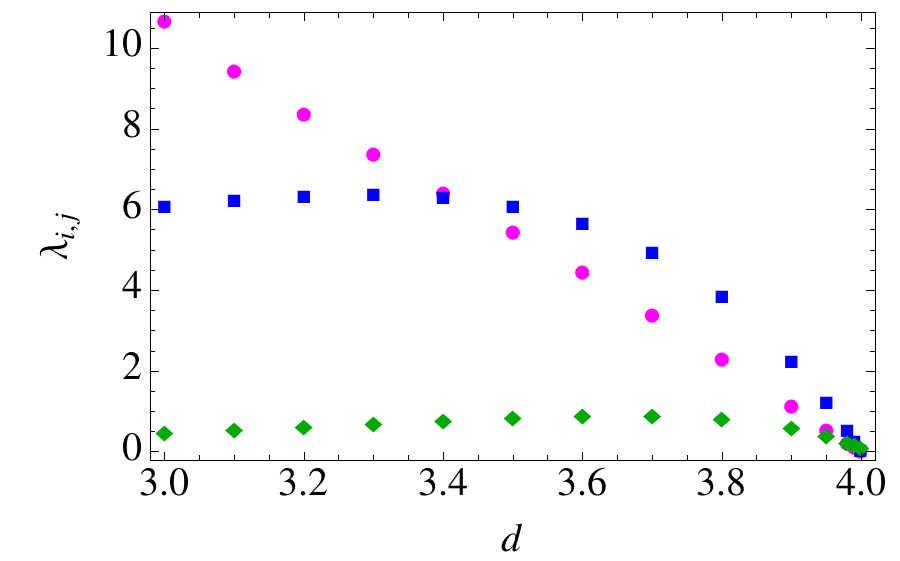}
\includegraphics[width=\linewidth]{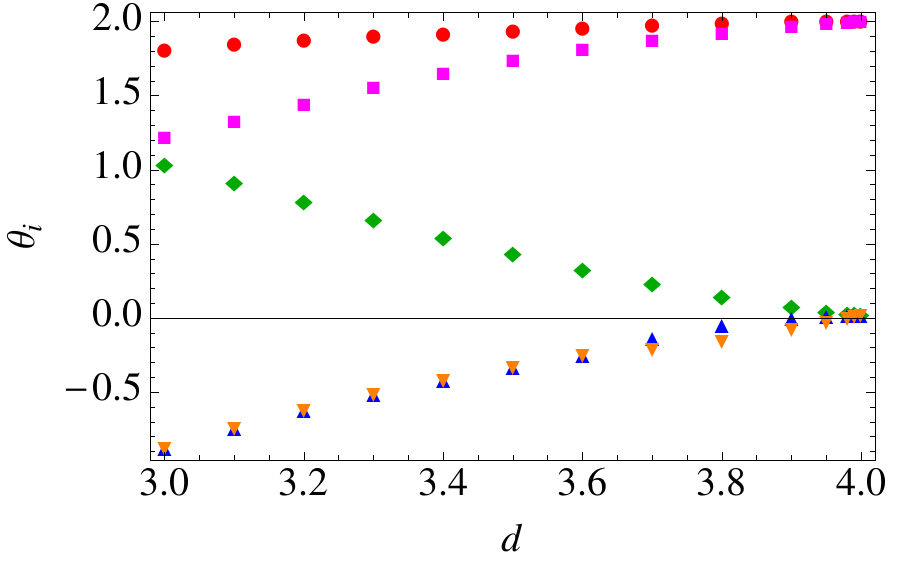}
\caption{\label{AFP_d_4}(Color online)  Here we show the fixed point values (upper panel) for $\lambda_{2,0}$ (magenta discs), $\lambda_{1,1}$ (blue squares) and $\lambda_{0,2}$ (green diamonds) at the AFP as a function of $d$ in LPA 4th order. Additionally, we show the real part of the five largest critical exponents (lower panel) that approach the asymptotic values $2$ and $0$, as expected from canonical scaling arguments.}
\end{figure}

For the DFP, the SFP as well as the IFP, the non-existence in $d=4$ follows from the lack of an interacting fixed point for $O(N)$ models. Thus it remains to study the fate of the AFP, which approaches the GFP towards $d=4$, see Fig.~\ref{AFP_d_4}.
%

\section{Conclusions}\label{conclusions}

In this paper we have established the nonperturbative functional Renormalization Group as a useful tool to study multicritical behavior in models with two competing order parameters, and address open questions in these models.
As our method relies on a truncation of the flowing action, we have tested the reliability of our truncation by confirming results on the isotropic, the decoupled and the biconical fixed point. We have also evaluated the crossover exponents $\phi_T$ as a function of $N$ at the isotropic fixed point, which are in good agreement with results from higher-order perturbative calculations.
Our analysis implies that the isotropic, bicritical fixed point is the stable one for $N_1= 1=N_2$. At $N_1=1$, $N_2=2$, and vice versa, this fixed point becomes unstable as it shows an additional relevant direction.  Our most sophisticated truncation shows that the decoupled tetracritical fixed point only becomes stable for values of $N=N_1+N_2 >3$.
We conclude that models with $N_1=1, N_2=2$, describing, e.g., anisotropic antiferromagnets in an external magnetic field, show tetracritical behavior, described by the biconical fixed point. Consequently we would expect these physical systems to exhibit a mixed phase, e.g., a supersolid in the case of Helium 4 or a biconical phase for anisotropic antiferromagnets, if the microscopic parameters of the material correspond to the $\Delta'>0$ region.
Here, we analyze the stability region of the biconical fixed point in 8th order LPA. We observe that this fixed point becomes stable as soon as the IFP becomes unstable. For larger values of $N$, the BFP then exchanges stability with the DFP.

We further elucidate that the system is dominated by collisions of fixed points, moving through coupling space as a function of $N_1, N_2$. As they collide, they exchange stability properties. Starting from a $\phi \leftrightarrow \chi$ symmetric fixed point at $N_1=1=N_2$ characterized by $\Delta'<0$ and thus bicritical behavior, this fixed point starts to move through coupling space, leaving the $\phi \leftrightarrow \chi$ symmetric regime, as we increase $N_2$. 
At $N_{2\, c}$, it collides with the IFP, and breaks through the $\Delta'=0$ surface. It then becomes the biconical, stable fixed point describing tetracritical behavior. At a second critical value of $N_2$ the BFP collides with the DFP. Again, these two interchange their stability properties in the collision, with the DFP becoming stable.

This observation, consistent with Ref.~\onlinecite{Moser2008}, could explain why the critical exponents at the BFP in the $N_1=1$, $N_2=2$ case will be rather close to that of the Heisenberg universality class, if $N_c \lesssim 3$. If these fixed points collide very close to that point, their critical exponents are still very close together at the physical point $N_1=1, N_2=2$.  
This analysis of the BFP also shows an interesting connection to the cubic fixed point\cite{Aharony2,Bornholdt:1994rf,Bornholdt:1995rn,Delamotte2002}: The $\phi \leftrightarrow \chi$ symmetric cubic fixed point coincides with the biconical fixed point for $N_1=1=N_2$.

Further, we report the discovery of a new, asymmetric fixed point at $\Delta'<0$ within our truncation. Whether this fixed point persists also beyond our approximation remains to be confirmed. In the affirmative case, the $\Delta'<0$ region of the theory space could be dominated by a bicritical fixed point. Thus models with initial conditions in this region could also show bicritical behavior.

Turning to $d=4$, we study several implications for high-energy physics, with potential implications  for hybrid models for inflation as well as models with a Higgs-inflaton coupling.
Within our truncation we confirm that these models suffer from a triviality problem, as none of the  three-dimensional fixed points has a non-trivial continuation at $d=4$. 

We further use the example of the semi-Gau\ss{}ian fixed points to argue that UV complete models in which a sector of the theory becomes non-interacting, can still show non-trivial critical exponents associated with this sector. Such a behavior could be of interest for UV completions of the Standard Model coupled to gravity.

Finally, we discuss how Goldstone modes can emerge from models with only discrete symmetries. A necessary requirement is the existence of symmetry enhanced points in theory space, such as, in our case the $Z_2 \oplus Z_2 \rightarrow O(2)$ point. We speculate, whether infrared-attractive symmetry enhanced points could help to construct mass hierarchies in scalar models.

To summarize, we have established the FRG as a worthwhile tool to investigate models with multiple  order parameters. In the future, several exciting extensions of our study are possible:
Following the methods developed in extended studies of $O(N)$ models should allow us to improve our estimate for the anomalous dimension in order to achieve quantitative precision. Furthermore, the FRG naturally lends itself to (numerical) studies of the non-universal flow of the effective potential towards first-order phase transitions, and the study of full phase diagrams, e.g., for the case of anisotropic antiferromagnets, as well as the $SO(5)$ theory of high-$T_c$ superconductors and systems including fermionic degrees of freedom.  As we have derived the flow equations for general dimensions $d$, and performed no expansion around any value of $d$, an extension towards the highly interesting case of $d=2$ is possible, similar to Refs.~\onlinecite{Morris:1994jc,Codello:2012ec}. Finally, the one-loop form of the effective action makes it feasible to study models with more than two competing orders, and discuss the case of multicritical points with a higher multiplicity than four.\newline\\

{\bf{\it{Acknowledgements}}}\\
We gratefully acknowledge discussions with J.~Berges, J.~M.~Pawlowski,  C.~Wetterich, S.~Wetzel  and useful correspondence with M.~Hasenbusch and A.~Pelissetto. We thank H.~Gies for helpful comments on the manuscript. M.M.S. thanks the Perimeter Institute for hospitality during a part of this work.
Research at Perimeter Institute is supported by the Government of Canada through Industry Canada and by the Province of Ontario through the Ministry of Research and Innovation. D.M. is supported by the Deutsche Forschungsgemeinschaft within the SFB 634. M.M.S. is supported by the grant ERC- AdG-290623. 


\end{document}